\documentstyle[aps,amstex,amssymb,graphics,preprint,epsf]{revtex}
\begin{document}                                                       

\draft 


\title {Hydrodynamic fluctuation forces}

\author{B.I.\ Ivlev}
 
\address{Department of Physics and Astronomy\\
University of South Carolina, Columbia, SC 29208\\
and\\
Instituto de F\'{\i}sica, Universidad Aut\'onoma de San Luis Potos\'{\i}\\
San Luis Potos\'{\i}, S. L. P. 78000 Mexico}   

      
\maketitle

\begin{abstract}

Two interaction mechanisms of particles in a fluid are proposed on base of forces, mediated by hydrodynamic thermal 
fluctuations. The first one is similar to the conventional van der Waals interaction, but instead of been mediated by 
electromagnetic fluctuations, it is mediated by fluctuations of hydrodynamic sound waves. The second one is due to a 
thermal drift of particles to the region with a bigger effective mass, which is formed by the involved surrounding 
fluid and depends on an inter-particle distance. The both mechanisms likely can be relevant in interpretation of the 
observed long-range attraction of colloidal particles, since a set of different experiments shows the attraction energy of
the order of $k_{B}T$ and, perhaps, only a fluctuation mechanism of attraction provides this universality. 
\end{abstract}
\vspace{0.7cm} 


\narrowtext
\section{INTRODUCTION}
Charge-stabilized colloidal suspensions exhibit a variety of unusual physical properties \cite{DER,ISR,RAJ}. Colloidal 
particles can be organized into crystal \cite{PIER} and into structures with clusters and voids 
\cite{ONODA,ISE,JAIME1,JAIME2,JAIME3,JAIME4}. A system of colloidal particles may undergo different types of phase 
 transitions \cite{PINCUS,SINHA,GAST,LARSEN,RADZ,LIN,SOW}. The topological phase transitions in two dimensional systems of 
colloidal particles have been reported in Refs.~\cite{MURRAY,ZAHN}. In Ref.~\cite{NEL} buckling instabilities in a confined 
colloidal crystals were analyzed. An interesting behavior of colloids in external fields were reported in 
Ref.~\cite{RUBI1}. Colloidal particles accept in an electrolyte some effective charge screened by counterions at Debye's 
length $\lambda_{D}$, what is described by the repulsion potential of Derjaguin, Landau, Verwey, and Overbeek (DLVO)
\cite{DER,ISR}. The DLVO theory, as a result of solution of the linearized Poisson-Boltzmann equation, has been questioned
in Refs.~\cite{SINHA,MADDEN}. The generalization of DLVO interaction via a modification of counterion screening was 
reported in Ref.~\cite{OSPECK}. 

Despite a long history of the problem, the interaction of colloidal particles remains a matter of challenge and 
controversy. The authors of Ref.~\cite{SINHA}, studying phase transitions in charged colloidal systems, found a substantial
deviation from predictions resulting from a screened Coulomb interaction. Experimental data \cite{TATA} suggest a net 
attraction of particles for explanation of measurements (see also the Comment on that work \cite{PALBERG}). Later an 
attraction of micron size particles separated by a micron-sized distance has been established experimentally. The
interaction potential has been found to have a minimum $-U_{0}$ at center-to-center particle distance $R_{0}$ of the micron 
size. In the work \cite{FRAD} colloidal particles, confined between two glass planes, exhibit $U_{0}\simeq 0.2k_{B}T$ for 
different ionic strengths. This type of experimental arrangement, studied in another lab \cite{ARAUZ}, gives 
$U_{0}\simeq (0.3 - 0.4)k_{B}T$ for different particles diameter and distances between planes. Another study of colloidal 
particles, confined between two glass planes, gives $U_{0}\simeq 1.3k_{B}T$ or less for different particle diameters and 
distances between planes \cite{GRIER1,GRIER2} (see also \cite{GRIER3,GRIER4}). The attraction of $0.5k_{B}T$ per neighbor 
particles (separated by the micron-sized distance) has been deduced for the colloidal crystal in Ref.~\cite{GRIER3}. 
Polystyrene colloidal particles of a diameter $0.5\mu{\rm m}$ on the water-air interface exhibit $U_{0}\simeq 0.5k_{B}T$ at 
$R_{0}\simeq 0.9\mu{\rm m}$ \cite{JAIME5}. For micron sized colloidal ``molecules'' on the air-water interface 
\cite{JAIME1,JAIME2,JAIME3} the binding energy can be estimated as a few $kT$. For bound particles on the fluid-fluid 
interface an approximate estimation is $U_{0}\sim 4kT$ \cite{NIKOL}. As one can see, despite of different conditions 
(even particles on interfaces) there is a very stable common feature of all various experiments: the attraction minimum is 
always of the order of $k_{B}T$. This provides a hypothesis of some common mechanism of the micron-sized attraction, which 
is responsible for the universality of $U_{0}\sim k_{B}T$. Note, mean-field energies in colloidal physics (electrostatic and
hydrodynamic) are a few orders of magnitude bigger comparing to $kT$ at room temperature.

Let us analyze some mechanisms of attraction proposed in the literature. 

A principal question is that can like-charged particles separated by a micron distance attract each other due to a solely 
electrostatic mean-field interaction (for example, by some charge redistribution) which is not accounted by DLVO theory. 
This type of attraction was predicted in Ref.~\cite{SOGAMI}. The results of Ref.~\cite{SOGAMI} are not applicable to 
dielectric particles but only to those containing inside the electrolyte identical to outside one. The correct calculation
of the interaction of that kind shows only repulsion in contrast to Ref.~\cite{SOGAMI}. The electrostatic attraction 
between like-charged particles based on the mean-field approach was also predicted in Ref. \cite{BOWEN}, but that 
conclusion was incorrect, as shown in the works \cite{NEU,SADER}. An attraction potential force of like-charged particles 
at a micron distance due to a mean-field mechanism seems to be extremely unlikely and the works \cite{NEU,SADER} provide 
strong arguments in support of this statement.

The conventional van der Waals attraction, mediated by the high-frequency (visible light) electromagnetic fluctuations, is
very small in the micron range, consisting less than $10^{-2}T$ at room temperature \cite{STAT,NIN1,NIN2,NIN3}, and cannot 
provide the observed attraction. Measurements of this type of attraction in colloidal systems at short distances of the 
order of a few hundred Angstroms have been performed in Refs.~\cite{SURESH,BEVAN}. An analysis of attraction at such short 
distances, including influence of the surface roughness, is given in Refs.~\cite{KIEFER,CZARNECKI1,CZARNECKI2,PRIEVE}.

The van der Waals interaction mediated by the low frequency (order of the plasma frequency) electromagnetic fluctuations 
decays fast with a distance $R$ between two particles as $-V_{0}\exp(-2R/\lambda_{D})$ \cite{AU,NIN2}. $V_{0}$ has in our case the
thermal fluctuation origin and is less than strong electrostatic repulsion energy even at $R\simeq\lambda_{D}$. The 
exponential dependence $\exp(-R/\lambda_{D})$ of the attraction potential in Ref.~\cite{TOKU}, dealing with the same effect, 
differs from the above correct exponent \cite{AU,NIN2}. The results \cite{TOKU} cannot explain the attraction at 
$R\sim 10\lambda_{D}$ in Ref.~\cite{GRIER1} at least for the reason the attraction, reported in \cite{TOKU}, is too small on 
that distance. 

The effect of correlation of counterions \cite{MANN,BRUIN,LOWEN,SHKL1,SHKL2} (see also \cite{LAU,MESSINA,TRIZAC}) results in
their rearrangement in the vicinity of a colloidal particle of the order of mean distance between counterions, which is 
less than $\lambda_{D}$. On that short distance an attraction is possible as shown in 
Refs.~\cite{MANN,BRUIN,LOWEN,SHKL1,SHKL2}. From a position on the long micron-sized distance, the correlation of 
counterions forms the effective charge $Ze$ of the particle, which is screened at $\lambda_{D}$ and determines the repulsive 
part of the interaction. 

Depletion forces between colloidal particles originate from an influence of the thermodynamic energy by the finite size of
small particles, which constitute the surrounding medium \cite{PARKER,KAPLAN,GOET,VERMA,METTEO}. The range of the depletion
interaction is determined  by the size of small particles, which can be not necessary ``small'' like in Ref. \cite{VERMA},
where surrounding polymer coils have the macro-particle size. When the surrounding electrolyte contains only microions, 
the depletion force is of a very short range comparing to the micron scale.

In Ref. \cite{MUD1} a mechanical effect is proposed for explanation of the observed motion of particles towards each other,
when they move away a single wall due to the Coulomb repulsion \cite{GRIER3,GRIER4}. This mechanical effect is irrelevant 
in analysis of interaction of free particles in a fluid or ones confined between walls or on interfaces (see also 
\cite{COMM}).

As one can conclude, no one above mechanism is responsible for the set of observations of long distance attraction of 
colloidal particles (may be excepting the mechanical effect \cite{MUD1}, which can be relevant for the particle motion in 
the one wall geometry \cite{GRIER3}). An important hint in search and selection of possible mechanisms of attraction is the 
universality of a depth of the attraction minimum $U_{0}\sim k_{B}T$. A most probable mechanism, satisfying this criterion, 
is an interaction, mediated by thermal fluctuations of some physical values. This is a microscopic type of interaction, 
which can be formulated in terms of a potential energy. Regardless of specificity of fluctuating matter, the free energy 
of thermal fluctuations is always proportional to $k_{B}T$. For example, the attraction potential of two dielectric 
particles (refractive index $n$) in the water (refractive index $n_{0}$) due to thermal fluctuations of the 
electromagnetic field is proportional to $c(n,n_{0})k_{B}T$. If $n$ is close to $n_{0}$, then $c\sim (n-n_{0})^{2}$, according to 
the perturbation theory. With real values of $n$ and $n_{0}$ for polystyrene and water the coefficient $c$ is very small, 
what makes the conventional van der Waals attraction in the micron range to be negligible. Nevertheless, if to put 
formally $n\rightarrow\infty$ (zero fields gradient on the particles surface), the attraction in the micron range is not small
comparing to $k_{B}T$ \cite{STAT}. This is similar to the situation in hydrodynamics, where fluctuating electromagnetic 
waves are substituted by hydrodynamic ones and the fluid velocity is zero on the particles surface. Hence, one can expect 
the interaction, mediated by fluctuating hydrodynamic waves, to be not small like in the electromagnetic case with 
$n\rightarrow\infty$.

The nature of forces, mediated by fluctuations of sound waves in a fluid, can be clarified in the following way: the 
energy of thermal fluctuations of the fluid depends on a distance between two particles, which play a role of obstacles 
for a fluid motion, and hence this results in a force. The non-electromagnetic type of fluctuation forces, discussed by 
Dzyaloshinskii, Lifshitz, and Pitaevskii \cite{DZYAL}, can be considered as some sort of the van der Waals \cite{STAT} or 
the Casimir \cite{CAS} forces. We do not concern here the story of origin of the two titles for fluctuation interactions. 

In addition to the attraction, mediated by fluid fluctuations, there is another mechanism of a fluctuation interaction 
related to hydrodynamics: this is an interaction, mediated by thermal fluctuations of particle positions in a fluid. The
effective particle masses depend on a fluid mass, involved into the motion. The fluid mass depends on an inter-particle
distance and therefore the effective particle masses also depend on that distance. If a mass of a classic non-dissipative 
harmonic oscillator depends on the coordinate, the mean coordinate shifts to the region with a bigger mass (smaller 
velocity), since the particle spends there more time. Analogously, for a Brownian motion of particles     in a fluid a thermal
drift occurs in the direction of a bigger effective mass.  As shown in the paper, the effective particle mass, which 
determines the effect, should be calculated on base of the Euler (non-dissipative) hydrodynamics and it can be called the
Euler mass. 

The above fluctuation interactions relate to hard spheres (no mean-field interaction excepting an infinite repulsion on 
contact). If particles are charged, the fluctuation hydrodynamic forces are weakly modified by the Coulomb effects in a 
bulk fluid under typical experimental conditions. The only interaction, mediated by fluctuations of particle positions, 
can be influenced by an electrostatic restriction of particle fluctuation motions in the direction perpendicular to 
charged confining plates. The total interaction is a sum of the fluctuation ones and the Coulomb repulsion.

The following results are presented in the paper: (i) the attraction forces between two parallel plates and spherical 
particles with short inter-surface distance, mediated by hydrodynamic fluctuations of sound waves, are shown to exist and 
they are calculated analytically; (ii) a novel interaction of particles in a fluid is proposed, which is based on 
dependence of their effective masses on a distance between them; (iii) the Fokker-Planck equation for two particles in a 
fluid is derived; (iv) a difference in measurements of a particle interaction by optical tweezers and by a long time 
statistics is pointed out and calculated. 
\section{THE HYDRODYNAMIC VAN DER WAALS INTERACTION} 
Suppose two particles are placed inside a hydrodynamic medium, they are totally fixed in space, and serves only as 
obstacles for a fluid motion. There is no macroscopic motion in the system and the only motion is caused by thermal 
fluctuations of the fluid velocity $\vec v(\vec r,t)$. In this case the free energy of thermal fluctuations of the fluid 
$F(R)$ depends on the distance $R$ between bodies. The function
\begin{equation}
\label{1}
U_{vdW}(R)=F(R)-F(\infty)
\end{equation}
is an interaction mediated by fluid fluctuations. Analogously to the conventional van der Waals interaction mediated by
electromagnetic fluctuations, the potential (\ref{1}) can be called {\it the hydrodynamic van der Waals interaction}. 
To find the free energy of thermal fluctuations of the fluid one can start with the linearized Navier-Stokes equation 
\cite{HYDR}
\begin{equation}
\label{e1}
\rho\frac{\partial\vec v}{\partial t}=-\nabla p + \eta\nabla^{2}\vec v + \left(\zeta + \frac{\eta}{3}\right)\nabla{\rm div}\vec v
\end{equation}
There are two types of fluid motion, one of them is a transverse diffusion and the second one is longitudinal sound waves, 
associated with the density variation. The equilibrium free energy of transverse motions is determined by the Boltzmann
distribution of their kinetic energies and does not depend on the friction coefficient in the thermal limit. Since there 
is no static interaction for transverse motions, their equilibrium free energy in the thermal limit depends on the total 
volume, but not on relative positions of bodies. Therefore, transverse fluctuations do not result in an interaction. Quite
opposite situation occurs for longitudinal motions, when the total free energy is a sum of energies of different sound 
modes. The spectrum of sound waves depends on the distance between bodies $R$ due to hydrodynamic boundary conditions on 
body surfaces and this results in $R$-dependence of the free energy. Hence, the fluctuation interaction between bodies is 
mediated by hydrodynamic sound waves like the conventional van der Waals interaction is mediated by fluctuations of 
electromagnetic ones. Putting $\vec v=\nabla\partial\phi/\partial t$, one can obtain from Eq.~(\ref{e1})
\begin{equation}
\label{e2}
\rho\hspace{0.1cm}\frac{{\partial}^{2}\phi}{\partial t^{2}}=-\delta p + 
\left(\zeta + \frac{4\eta}{3}\right)\frac{\partial}{\partial t}\nabla^{2}\phi
\end{equation}
Through thermodynamic relations and the continuity equation one can obtain $\delta p=-\rho s^{2}_{0}{\nabla}^{2}\phi$, where 
$s_{0}$ is the adiabatic sound velocity \cite{HYDR}. At the typical frequency $\omega\sim s_{0}/a$ ($a$ is the particle radius),
involved into the problem, the dissipative term in Eq.~(\ref{e2}) is small and one can write
\begin{equation}
\label{e3}
\frac{{\partial}^{2}\phi}{\partial t^{2}}-s^{2}_{0}{\nabla}^{2}\phi = 0
\end{equation}
According to the case of a small friction, the boundary condition ${\nabla}_{n}\phi=0$ to Eq. (\ref{e3}) corresponds to the 
Euler equation \cite{HYDR}. From general point of view, free energy of a system of harmonic oscillators does not depend on 
friction in the thermal limit.

Let us consider first the case of two infinite parallel plates, separated by the distance $R$, when the frequency spectrum
has the form \cite{HYDR}.
\begin{equation}
\label{e4}
\omega{^2}_{n}(k) = s^{2}_{0}\left(k^2 + \frac{\pi^2 n^2 }{R^2}\right)
\end{equation}
The free energy per unit area of the system now can be expressed as a sum of energies of independent oscillators
\begin{equation}
\label{e5}
F=T\int\frac{d^2 k}{(2\pi)^2}\sum^{\infty}_{n=1}\ln\frac{\hbar\omega_{n}(k)}{T}
\end{equation}
This type of interaction has been considered in literature \cite{NIN1}. The simplest way to calculate the energy (\ref{e5})
is to divide the whole interval $R$ by small segments $a_{0}=R/N$. Then
\begin{equation}
\label{e6}
\sum^{N}_{n=1}\ln\left(k^2 + \frac{\pi^2 n^2}{a^{2}_{0}N^2}\right)=2\sum^{N}_{n=1}\ln\frac{n}{N}+
\ln\prod^{\infty}_{n=1}\left(1+\frac{a^{2}_{0}N^2 k^2}{\pi^2 n^2}\right)
\end{equation}
Using the relation $x\prod^{\infty}_{n=1}(1+x^{2}/\pi^{2}n^{2})=\sinh x$, the Stirling formula for $N!$, omitting constants, and 
$R$-linear terms, one can obtain from Eqs. (\ref{e4}-\ref{e6})
\begin{equation}
\label{e7}
F=\frac{T}{2}\int\frac{d^2 k}{(2\pi)^2}\ln\left[1-\exp(-2kNa_{0})\right]
\end{equation}
Now one should put in Eq. (\ref{e7}) $Na_{0}=R$ and after integration we obtain, according to Eq. (\ref{1}), the van der Waals
interaction per unit area of two infinite plates
\begin{equation}
\label{e8}
u_{vdW}(R)=-\frac{\zeta(3)}{16\pi}\hspace{0.1cm}\frac{T}{R^2}
\end{equation}
Despite the van der Waals interaction (\ref{e8}) formally coincides with the electromagnetic van der Waals formula at
$\varepsilon\rightarrow\infty$ \cite{DZYAL}, existence of such effect in hydrodynamics is not trivial as follows from Section 
III. 

The result (\ref{e8}) enables to calculate the van der Waals energy $U_{vdW}(R)$ of two spheres of radii $a$, separated by 
the center-to-center distance $R$, when $(R-2a)\ll a$. In this case the inter-surface distance, measured in the direction 
parallel to the center-to-center line, is $R-2a +(x^2 +y^2)/a$, where $x$ and $y$ are the coordinates in the plane 
perpendicular to the center-to-center line. The procedure turns out to an integration
\begin{equation}
\label{e9}
U_{vdW}(R)=\int dxdy\hspace{0.1cm}u_{vdW}\left(R-2a+\frac{x^2 +y^2}{a}\right)
\end{equation}
The result of integration at $(R-2a)\ll 2a$ is
\begin{equation} 
\label{e10}
U_{vdW}(R)=-\frac{\zeta(3)}{16}\hspace{0.1cm}T\frac{a}{R-2a}\hspace{1.5cm}({\rm spheres})
\end{equation}
The interaction energy (\ref{e10}) is calculated under fixed boundaries of particles, which corresponds to zero sound
velocity $s$ of the particles material (infinite acoustic mismatch). Under reduction of the mismatch $U_{vdW}$ decreases,
turning to zero at $s=s_{0}$, if there is only longitudinal acoustic mode inside particles. Sound velocities for 
polystyrene particles $s\simeq 2.1\times 10^{5}{\rm cm}/{\rm s}$ and for water $s_{0}\simeq 1.5\times 10^{5}{\rm cm}/{\rm s}$ provide a 
finite mismatch, which reduces the result (\ref{e10}), as one can show, approximately four times. But in reality this 
conclusion is not correct, since transverse acoustic modes of the particle material increase the mismatch turning the 
interaction towards the result (\ref{e10}). An account of the finite mismatch is a matter of further study, but the exact 
$U_{vdW}$ seems to be close to the result (\ref{e10}).

The above calculations are applicable, strongly speaking, to uncharged particles like hard spheres. For charged particles
an electric charge density of a fluid $en$ is finite and decays on the Debye length $\lambda_{D}$. It results in a 
modification of the spectrum of fluctuations by adding the plasma frequency $s^{2}k^{2}+\omega^{2}_{p}n/n_{0}$, where 
$\omega^{2}_{p}=4\pi ne^{2}/\varepsilon M$ ($M$ is the mass of the fluid molecule) and $n_{0}\sim 10^{23}{\rm cm}^{-3}$ is the molecular
density of the fluid. The coefficient $n/n_{0}$ comes from that the sound spectrum is formed by the all fluid, but the 
Coulomb gap is determined only by a small amount of ions $n\sim Z/4\pi a\lambda^{2}_{D}\sim 10^{17}{\rm cm}^{-3}$ (a typical 
$Z\sim 10^{4}$). Since the wave vector $k$ is inversely proportional to the inter-particle distance, one can conclude from
here that the Coulomb effects modify the above hard sphere result (\ref{e10}), when $(R-2a)$ exceeds $100\mu{\rm m}$. 
\section{THE PARADOX}
In case of the conventional van der Waals interaction, mediated by electromagnetic fluctuations, the mean values of 
electric and magnetic fields are zero $<\vec E>=<\vec H>=0$. The stress tensor $\sigma_{ik}$ for electromagnetic field is 
quadratic with respect to fields and hence the mean value $<\sigma_{ik}>$ is not zero. This makes the origin of the force due
to electromagnetic fluctuations to be straightforward. The situation with hydrodynamic fluctuation forces is different. 
For the linearized Navier-Stokes equation the mean value of the velocity is zero $<\vec v>=0$ (the same for a fluctuation 
part of $p$). The hydrodynamic stress tensor
\begin{equation}
\label{11}
\sigma_{ik}=\eta\left(\frac{\partial v_{i}}{\partial r_{k}}+\frac{\partial v_{k}}{\partial r_{i}}\right)-p\delta_{ik}
\end{equation}
is linear in fluctuation variables, its fluctuation part is zero, and a fluctuation force has to be zero in this 
approximation. A non-zero contribution to $<\sigma_{ik}>$ can result from the nonlinear terms in the Navier-Stokes equation 
neglected in the above approach. Account of this non-linearity has been done in Ref.~\cite{MUD2} and a finite fluctuation 
force has been obtained. This result was shown to be incorrect in Ref.~\cite{MUD3}, where the exact mean value of the 
stress tensor (\ref{11}) was found to be zero on the base of exact non-linearity of the Navier-Stokes equations. A 
conclusion of Ref.~\cite{MUD3} is that a hydrodynamic fluctuation interaction is impossible, which contrasts with the result
(\ref{e8}). What is on a wrong track?

\begin{figure}[p]
\begin{center}
\vspace{1.5cm}
\epsfxsize=\hsize
\epsfxsize=13cm
\leavevmode
\epsfbox{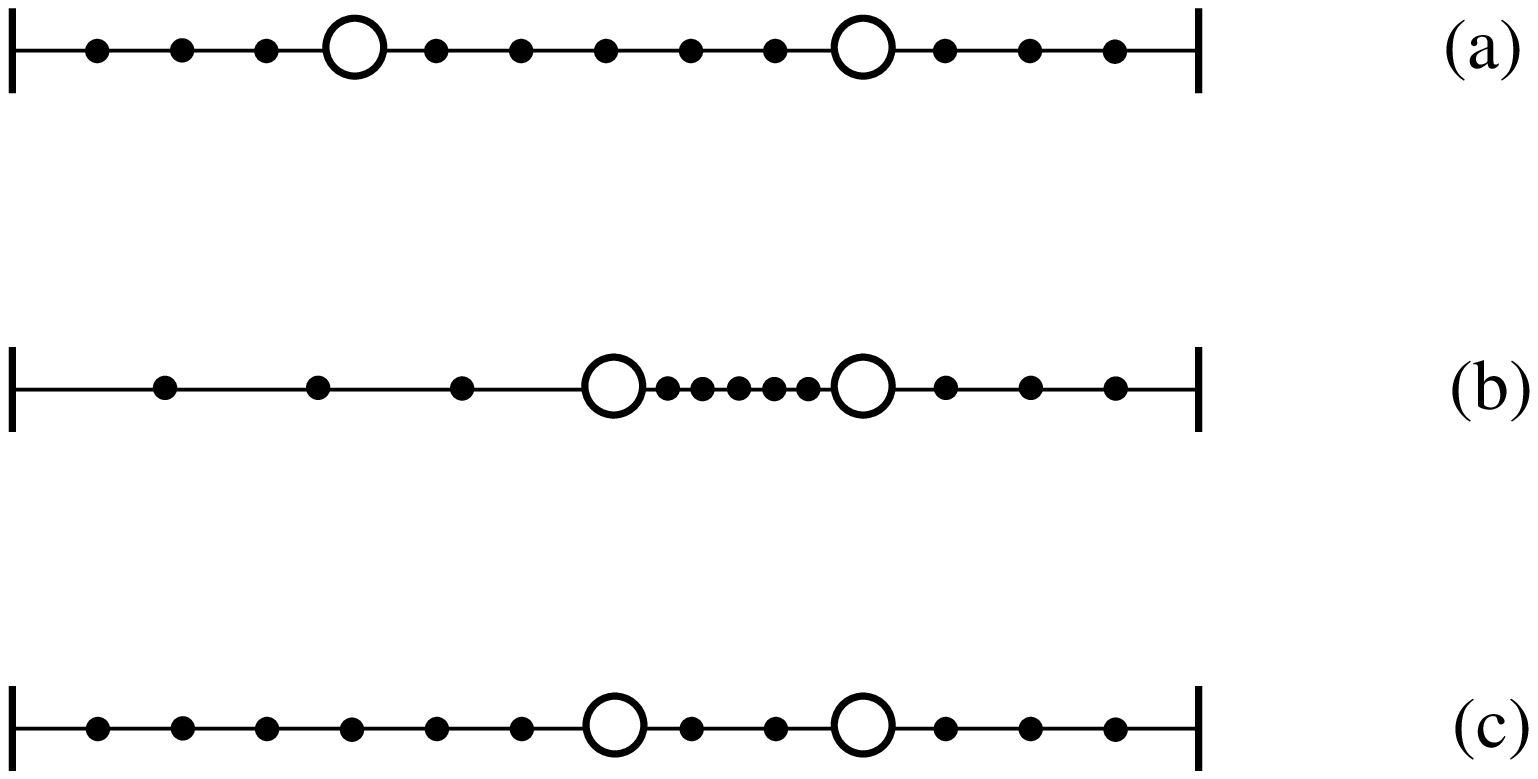}
\vspace{2cm}
\caption{The linear chain of particles. Each horizontal segment between particles behaves like an elastic harmonic spring.
The configuration (a) represents the ``natural'' positions of two attached particles (open circles). The configuration (b)
is obtained from (a) by a compression motion of attached particles without a destruction of harmonic bonds. (c) is another
``natural'' position obtained from (a) by a destruction of harmonic bonds.}
\label{fig1}
\end{center}
\end{figure}

To understand the situation let us consider the linear chain of small  particles, connected by elastic springs, shown in 
Fig.~\ref{fig1} and described by the dynamic equation
\begin{equation}
\label{12}
\frac{\partial^{2}u_{n}}{\partial t^{2}}=\frac{s^2}{b^{2}_{0}}(u_{n+1}+u_{n-1}-2u_{n})
\end{equation} 
where $s$ is the sound velocity and $b_0$ is the period. Two no-moved big particles substitute small particles, as shown in
Fig.~\ref{fig1}. The system is elastic and the force acting on a big particle, placed on the site $n$ is
\begin{equation}
\label{13}
F_{n}=\frac{ms^2}{b^{2}_{0}}(u_{n+1}-u_{n-1})
\end{equation}   
Here $m$ is the mass of a a small particle. The free energy of fluctuation motion of small particles $U_{vdW}$ is determined
by summation on self-frequencies $\omega_{i}$ of the system $F=T\sum\ln(\hbar\omega_{i}/T)$, resulting in
\begin{equation}
\label{14}
U_{vdW}=\frac{T}{2}\ln 4N
\end{equation}
where $N$ is a number of springs between two big particles. Two different positions of big particles, ``natural'' in 
Fig.~\ref{fig1}(a) and ``compressed'' in Fig.~\ref{fig1}(b), have identical self frequencies since the system is harmonic 
and hence $U^{(a)}_{vdW}=U^{(b)}_{vdW}$ (equal $N$ in Eq.~(\ref{14})). In this situation there is no van der Waals force, what is
clear, since the mean value of the linear force (\ref{13}) should be zero. On the other hand, for another ``natural'' 
position in Fig.~\ref{fig1}(c) self frequencies differ from those in Figs.~\ref{fig1}(a) and 1(b), in this case
$U^{(a)}_{vdW}\neq U^{(c)}_{vdW}$ (different $N$ in Eq.~(\ref{14})) and the van der Waals force is non-zero. It is a consequence 
of that a transition from the position (a) to the position (c) in Fig.~\ref{fig1} cannot occur within a harmonic 
approximation, one should destroy some harmonic springs and rearrange them again in a different way. The linear expression
for force (\ref{13}) is not valid to describe a transition from (a) to (c) and a real force is non-linear,  what makes its
average finite even for $<u_{n}>=0$.

An analogous situation takes place in hydrodynamics. According to its derivation, Eq.~(\ref{e8}) is valid only for 
discrete $R=Nb_{0}$, where $b_0$ is inter-atomic distance and $N$ is an integer number. The full dependence of $u_{vdW}$ 
on $R$ has a structure on the atomic scale corresponding to remove of discrete atomic layers from the inter-plane region.
The hydrodynamic expression for the stress tensor (\ref{11}) is not valid at such short scale, like Eq.~(\ref{13}) cannot 
describe breaking of harmonic bonds. In contrast to smooth van der Waals potential, mediated by electromagnetic 
fluctuations, the interaction, mediated by hydrodynamic ones, has a structure as a function of distance on the 
atomic scale, superimposed on the smooth function (\ref{e8}). To some extend, this is analogous to observation of the 
structured interaction potential \cite{METTEO}, where the role of atoms was played by small particles. The resulting 
statement is that the hydrodynamic expression for the stress tensor (\ref{11}) cannot be used for calculation of 
fluctuation forces since it becomes non-linear (and contributes to those forces) at short distances where the hydrodynamic
approach is not valid. Hydrodynamic fluctuation forces should be calculated on the base of energy like it is done in this
paper. The conclusion of Ref.~\cite{MUD3} on absence of hydrodynamic fluctuation forces based on use of the hydrodynamic 
stress tensor is incorrect.
\section{INTERACTION MEDIATED BY FLUCTUATIONS OF PARTICLE POSITION}
The hydrodynamic van der Waals interaction is formed by fast density fluctuations with a typical frequency 
$\omega_{L}\sim s_{0}/a\sim 10^{9}{\rm s}^{-1}$. Besides this longitudinal motion, there are also slow transverse fluctuations of the
fluid driven by fluctuations of particle linear velocities $\vec u_1$ and $\vec u_2$ of the small frequency 
$\omega_{T}\sim \eta/\rho a^2 \sim 10^{6}{\rm s}^{-1}$. This type of transverse fluctuations is independent of longitudinal one. 
Thermal fluctuations of particle velocities mediate an important part of the total interaction, which supplements 
$U_{vdW}$. 

Before consideration of this contribution we focus at first on some aspects of derivation of non-linear dissipative 
equations, which are useful to understand a formation of interaction by fluctuations of particle velocities. Let us start
with a one-dimensional motion in the potential $V(x)$ of a particle with the mass $m(x)$, attached to a heat bath, which 
provides a friction. A convenient way to proceed is to use the formalism of Caldeira and Leggett \cite{CALDLEG} of the 
infinite set of thermally equilibrium oscillators. The Lagrangian has the form
\begin{equation}
\label{e14}
L=\frac{m(x)}{2}\hspace{0.1cm}{\dot{x}}^{2}-V(x)+\frac{1}{2}\sum_{i}(m_{i}{\dot y}^{2}_{i}-m_{i}\omega^{2}_{i}y^{2}_{i})-
F(x)\sum_{i}c_{i}y_{i}
\end{equation}
where $F(x)=\int^{x}dz\sqrt{\eta(z)}$ is a non-linear coupling to the thermostat. Using the formalism of Caldeira and Leggett,
one can derive the Langevin equation in the limit of high temperatures
\begin{equation}
\label{e15}
m(x)\ddot{x}+\frac{1}{2}\hspace{0.1cm}\frac{\partial m}{\partial x}\hspace{0.1cm}{\dot{x}}^{2}+\eta(x)\dot{x}+V'(x)=\sqrt{\eta(x)}f(t) 
\end{equation}
where the average is defined as $\langle f(t)f(t')\rangle =2T\delta(t-t')$. Let us suppose the viscosity $\eta$ sufficiently big,
which separates the big frequency $\Omega\sim\eta/m$ from the low frequency $V''/\eta$ of the viscous motion in the potential
$V(x)$. One can also derive from the Lagrangian (\ref{e14}) the Fokker-Planck equation for the distribution function $W$, 
which in the low frequency limit depends only on $x$ and $t$
\begin{equation}
\label{e16}
\frac{\partial W(x,t)}{\partial t}=
\frac{\partial}{\partial x}\left\{\frac{W}{\eta(x)}\hspace{0.1cm}\frac{\partial}{\partial x}\left[V(x)+I(x)\right]+
\frac{T}{\eta(x)}\hspace{0.1cm}\frac{\partial W}{\partial x}\right\}
\end{equation}
The additional potential in Eq.~(\ref{e16}) is
\begin{equation}
\label{e17}
I(x)=-\frac{T}{2}\ln m(x)
\end{equation}
The Langevin (\ref{e15}) and the Fokker-Planck (\ref{e16}) equations are derived independently from the initial system
(\ref{e14}) and do not contain uncertainties since the high frequency limit is well defined by Eq.~(\ref{e15}). If to omit 
two mass terms in Eq.~(\ref{e15}), the high frequency limit becomes indefinite ($\Omega =\infty$) and an attempt to derive 
the Fokker-Planck equation from the Langevin equation meets the Ito-Stratonovich uncertainty \cite{ITOSTRAT} as a result of
loose of a definition of the high frequency limit. Note, neither the Ito nor the Stratonovich approaches result in the 
correct Fokker-Planck equation (\ref{e16}) for a massive particle with a variable viscosity $\eta(x)$. 

To clarify an origin of the effective potential (\ref{e17}) let us represent the variable $x$ in Eq.~(\ref{e15}) as 
$x(t)+\delta x(t)$, where the small correction $\delta x(t)$ varies rapidly with frequencies $\Omega\sim\eta/m$ and $x(t)$ is a slow
varying variable. In Eq.~(\ref{e15}) one can keep the second order of $\delta x$, considering $x(t)$ as an instant argument. 
After an average over high frequencies Eq.~(\ref{e15}) turns over into a low frequency part with two fluctuation induced 
terms
\begin{equation}
\label{e18}
\eta (x)\dot{x}+\left[\frac{\partial V(x)}{\partial x}-
\frac{1}{2}\hspace{0.1cm}\frac{\partial m(x)}{\partial x}\langle\delta\dot{x}^{2}\rangle-
\frac{1}{2\sqrt{\eta(x)}}\frac{\partial\eta(x)}{\partial x}\langle\delta xf\rangle\right]=\sqrt{\eta(x)}f(t)
\end{equation}
By means of the fluctuation-dissipation theorem \cite{STATPHYS1} one can write (see also \cite{BLATTER})
\begin{equation}
\label{e19}
\langle\delta\dot{x}^{2}\rangle =\frac{iT}{\pi}\int^{\infty}_{-\infty}\frac{d\omega}{m\omega+i\eta}=\frac{T}{m}
\end{equation}
The integration path in Eq.~(\ref{e19}) can be deformed into the far semi-circle at the upper half-plane of the complex 
$\omega$. Hence, $\langle\delta{\dot{x}}^{2}\rangle$ is determined by high frequencies and the use of instant variables, leading 
to incorporation of $\langle\delta{\dot{x}}^{2}\rangle$ into the low frequency equation (\ref{e18}), is correct. Analogously
\begin{equation}
\label{e20}
\langle\delta xf\rangle = -\frac{T\sqrt\eta}{\pi}\hspace{0.1cm}P\int^{\infty}_{-\infty}\frac{d\omega}{\omega(m\omega+i\eta)}=
\frac{T}{\sqrt\eta}
\end{equation}
The letter $P$ means the principal value of the integral, which can be represented as the integration along the infinite 
contour, consisting of the real axis plus a small circle around zero at the upper half-plane, and minus the integration 
along that small circle. The integration along the infinite contour can be shifted to the far upper half-plane and gives 
zero, but the integration along the small circle gives a finite result. Hence, $\langle\delta xf\rangle$ is determined by zero 
frequency limit, the use of instant argument and incorporation of $\langle\delta xf\rangle$ into Eq.~(\ref{e18}) are not
appropriate. Among two fluctuation induced term in Eq.~(\ref{e18}) only one (with $\partial m/\partial x$) has sense and it 
results in the potential (\ref{e17}). This remains true in a more general case, when the linearized form of the dynamic 
equation is
\begin{equation}
\label{e21}
(-m\omega^{2}-i\omega\eta(\omega)+V'')\delta x_{\omega}=0
\end{equation}
In this case the mass $m$ is strictly defined by the condition $\eta(\omega)/\omega\rightarrow 0$ at $\omega\rightarrow\infty$ and the 
potential (\ref{e17}) is determined by the high frequency limit according to Eq.~(\ref{e19}), when only a mass term plays a 
role. 

An origin of the potential (\ref{e17}) can also be also understood from the following non-rigorous arguments. One can write 
formally the free energy $F=-T\ln(\Delta p\Delta x/\hbar)$, where momentum fluctuations $(\Delta p)^{2}\sim mT$ are formed on a short
time scale $\Omega^{-1}$ and fluctuations of the coordinate $(\Delta x)^{2}\sim Tt/\eta$ are slow. In the expression
\begin{equation}
\label{e22}
F=-\frac{T}{2}{\ln m} +\frac{T}{2}\ln\frac{\hbar^{2}\eta}{T^{2}t}
\end{equation}
the first term originates from fast fluctuations of momentum and corresponds to Eq.~(\ref{e17}).

The interaction (\ref{e17}) has a simple interpretation. Suppose the classical non-dissipative particle with the variable 
mass $m(x)$ moves with the total energy $E$ in the harmonic potential $\alpha x^{2}$. For a variable mass the mean 
displacement $\langle x\rangle\neq 0$, since in the region with bigger mass the particle spends more time having a smaller 
velocity. One can easily show, when $m(x)$ varies slowly on the scale of the particle amplitude, $\langle x\rangle$ can be 
calculated putting the constant mass $m(0)$ and adding the potential $-(E/2)\ln m(x)$. In other words, a particle tends to be 
more time in a region with bigger mass. This conclusion remains correct for a dissipative case with fast fluctuations of 
velocity instead of harmonic oscillations, when the energy $E$ should be approximately substituted by the temperature $T$.
This correspond to the potential (\ref{e17}). 

One can easily generalize the method for multi-dimensional case. Suppose in the high frequency limit the kinetic energy 
has the form 
\begin{equation}
\label{e23}
K=\frac{1}{2}\hspace{0.1cm}m_{ij}(\vec R)\dot{R}_{i}\dot{R}_{j}\hspace{1cm}(\omega\rightarrow\infty)
\end{equation}
Then in the equation of motion, where only the kinetic part is kept,
\begin{equation}
\label{e24}
m_{ij}(\vec R)\ddot{R}_{j}+\left(\frac{\partial m_{ij}}{\partial R_{k}}-
\frac{1}{2}\hspace{0.1cm}\frac{\partial m_{kj}}{\partial R_{i}}\right)\dot{R}_{k}\dot{R}_{j}=F_{i}
\end{equation} 
one can consider the variables again as a sum of slow and fast parts $R_{i}(t)+\delta R_{i}(t)$. An average of the quadratic 
(with respect to $\delta\dot{R}$) part produces the fluctuation induced effective force
\begin{equation}
\label{e25}
F^{ef}_{i}=F_{i}+
\frac{1}{2}\hspace{0.1cm}\frac{\partial m_{kj}}{\partial R_{i}}\hspace{0.1cm}\langle\delta\dot{R}_{k}\delta\dot{R}_{j}\rangle
\end{equation}
With account of the average $\langle\delta\dot{R}_{k}\delta\dot{R}_{j}\rangle=Tm^{-1}_{kj}$ the effective force reads
\begin{equation}
\label{e26}
F^{ef}_{i}=F_{i}+\frac{T}{2}\hspace{0.05cm}m^{-1}_{kj}\hspace{0.05cm}\frac{\partial m_{kj}}{\partial R_{i}}=F_{i}+
\frac{\partial}{\partial R_{i}}\left[\frac{T}{2}\ln({\rm det}\hspace{0.05cm}m)\right]
\end{equation}
From here the generalization of Eq.~(\ref{e17}) follows
\begin{equation}
\label{e27}
I(\vec R)=-\frac{T}{2}\ln\left[{\rm det}\hspace{0.05cm}m(\vec R)\right]
\end{equation}
The additional potential energies (\ref{e17}) or (\ref{e27}) have a fluctuation origin and are mediated by fast fluctuations
of velocity with the typical frequency $\Omega\sim\eta/m$, which form, for each instant coordinate, the quasi equilibrium 
free energy. One can formulate a rule of calculation the effective fluctuation potential even for a system with a 
complicated dynamics: one has to find the kinetic energy (\ref{e23}) in the limit of high frequency and to insert the mass 
tensor into Eq.~(\ref{e27}). The method in the presented form is applicable to classical systems with the kinetic energy 
proportional to square of velocities. 
\section{TWO PARTICLES IN A FLUID AND THE EULER MASS}
Suppose there are two particles in a fluid of radii $a$ separated by the center-to-center distance $R$. If they perform an
oscillatory motion with a high frequency $\omega$, the fluid velocity obeys the Euler equation everywhere in a fluid 
excepting a thing layer of the thickness $\sim\sqrt{\eta/\rho\omega}$ close to the particle surfaces, where the full 
Navier-Stokes equation should be used \cite{HYDR}. Hence for finding the mass tensor (\ref{e23}) one has to solve the Euler 
equation with the boundary condition for a normal component of the fluid velocity. For this reason, the mass, 
corresponding to the high frequency limit of particle dynamics, can be called the Euler mass. For example, the Euler mass 
tensor of one particle is \cite{HYDR}
\begin{equation}  
\label{e28}
m_{ij}=\frac{2\pi}{3}a^{3}(2\rho_{0}+\rho)\delta_{ij}
\end{equation}
where $\rho_{0}$ is the mass density of the particle. See also Refs.~\cite{LUB1,LUB2}.

In the case of two particles in a bulk fluid the Euler mass cannot be calculated analytically for an arbitrary relation 
between $R$ and $a$. Nevertheless, there is a situation, when an analytical calculation of $I(R)$ in a full range is 
possible. This is a case of two particle confined between two parallel plates separated by the distance of the particle 
diameter, which corresponds to the current experiments \cite{ARAUZ1}. Let us suppose the particles to be of cylindrical 
shape with the axis of the length $2a$ perpendicular to the plates. In this case particle and fluid velocities are 
directed parallel to the plates and the problem becomes two dimensional. A velocity of the incompressible Euler fluid can 
be written as $\vec v=\nabla\phi$ with a boundary condition for the normal derivative $\partial\phi/\partial\vec n$ canceling the 
normal components of particle velocities. Since the scalar $\phi$ satisfies the Laplas equation $\nabla^{2}\phi =0$ in two 
dimensions, one can use a conformal transformation to convert the geometry into a planar one. If the centers of two 
particles localize at positions ${\rm Re}z=\pm R/2$, ${\rm Im}z=0$ at the complex $z$-plane, the conformal transformation
\begin{equation}
\label{e29}
z=\left(\frac{R^{2}}{4}-a^{2}\right)^{1/2}\coth\frac{w}{2}
\end{equation}
maps the two circles on two infinite parallel lines 
\begin{equation}
\label{e30}
{\rm Re}\hspace{0.1cm}w=\pm\ln\left(\frac{R}{2a}+\sqrt{\frac{R^{2}}{4a^{2}}-1}\right)
\end{equation}
at the complex $w$-plane. For the plane geometry the problem can be solved in a straightforward way. The mass tensor is 
easily diagonalyzed by two center-of-mass motions and two relative ones. Using the formula (\ref{e27}), one can obtain after
some calculations
\begin{align}
\label{31}
I(R)=&-T\ln\left[1-\frac{2(R^{2}-4a^{2})}{a^{2}(1+\rho_{0}/\rho)}\sum^{\infty}_{n=1}
\frac{n(R-\sqrt{R^{2}-4a^{4}})^{2n}}{(R+\sqrt{R^{2}-4a^{2}})^{2n}+(2a)^{2n}}\right]\nonumber\\
&-T\ln\left[1+\frac{2(R^{2}-4a^{2})}{a^{2}(1+\rho_{0}/\rho)}\sum^{\infty}_{n=1}
\frac{n(R-\sqrt{R^{2}-4a^{2}})^{2n}}{(R+\sqrt{R^{2}-4a^{2}})^{2n}-(2a)^{2n}}\right]
\end{align}
At $\rho_{0}=\rho$ in limiting cases one obtains
\begin{equation}
\label{32}
I(R)=T
\begin{cases}
-\ln(\pi^{4}/72)+(6/\pi^{2})\sqrt{(R-2a)/a}\hspace{0.1cm};&\left(R-2a\right)\ll 2a\\
-a^{4}/R^{4}\hspace{0.1cm};&2a\ll R
\end{cases}
\end{equation}
$I(R)$ is plotted in Fig.~\ref{fig2} by a dashed line. The van der Waals interaction for cylindrical particles of the length
$2a$ can be obtained from Eq.~(\ref{e8}) in the same manner like Eq.~(\ref{e9}). At $(R-2a)\ll 2a$ for two cylinders of the 
length $2a$
\begin{equation}
\label{33}
U_{vdW}=-\frac{\zeta(3)}{16}\hspace{0.1cm}T\left(\frac{a}{R-2a}\right)^{3/2}\hspace{1.5cm}({\rm cylinders})
\end{equation}
The total interaction $U_{vdW}(R)+I(R)$ is shown in Fig.~\ref{fig2} by the solid line, where the close particles limit 
(\ref{33}) is extrapolated up to the region $(R/2a-1)\lesssim 1$. The interaction $I(R)$ is not influenced by the Coulomb 
effects in a bulk fluid, since it is mediated by fluctuations, corresponding to an incompressible fluid. Nevertheless, in a
restricted geometry, for example between two charged plates, fluctuation motion in some direction can be restricted, which
can modify the interaction $I(R)$.
\section{THE FOKKER-PLANCK EQUATION FOR TWO PARTICLES IN A FLUID}
The low frequency dynamic equations of particles
\begin{equation}
\label{e31}
\zeta^{ij}_{0}(\vec R)u^{j}_{1,2}+\zeta^{ij}_{1}(\vec R)u^{j}_{2,1}=F^{i}_{1,2}
\end{equation}
is non-linear due to a coordinate dependence of friction coefficients in Eq.~(\ref{e31}). The $\zeta_{1}$-term in Eq.~\ref{e31}
results in a macroscopic force on the particle (1) due to a macroscopic motion of the particle (2). This is called the 
hydrodynamic interaction \cite{MAZUR,LADD1,LADD2,LADD3}. 

The Fokker-Planck equation for the distribution function $W(\vec R,t)$ of two particles in a fluid can be derived in a 
manner like one leading to Eq.~(\ref{e16}). The result is the following
\begin{equation}
\label{e32}
\frac{\partial W}{\partial t}=\frac{\partial}{\partial R_{i}}\zeta^{-1}_{ij}\left[W\frac{\partial U_{tot}(R)}{\partial R_{j}}+
T\frac{\partial W}{\partial R_{j}}\right]
\end{equation}
where $\zeta_{ij}=(\zeta^{ij}_{0}-\zeta^{ij}_{1})/2$ and the total interaction potential consists of some mean-field electrostatic 
part $U(R)$ and two fluctuation interactions 
\begin{equation}
\label{e33}
U_{tot}(R)=U(R)+U_{vdW}(R)+I(R)
\end{equation}
The friction tensor in the limit $a\ll R$ has the form \cite{MAZUR}
\begin{equation}
\label{e34}
\zeta_{ij}=3\pi a\left[\left(1+\frac{3a}{4R}+\frac{9a^{2}}{16R^{2}}\right)\delta_{ij}+
\left(\frac{3a}{4R}+\frac{27a^{2}}{16R^{2}}\right)\frac{R_{i}R_{j}}{R^{2}}\right]
\end{equation}
For colloids experiments, which deal with long time statistics \cite{FRAD,ARAUZ}, the equilibrium distribution function
$W\sim\exp(-U_{tot}(R)/T)$ is relevant. In the optical tweezers experiments \cite{GRIER1,GRIER2,GRIER3,GRIER4} two particles are
initially fixed and statistics of initial motions after release is studied. If friction coefficients $\zeta^{ij}_{0,1}$ would
be coordinate independent, the both methods give the same interaction potential. But situation becomes different for 
coordinate dependent friction coefficients. If at the moment $t=0$ the distribution function was artificially localized by 
optical tweezers at the point $\vec R_{0}$  $(W=\delta(\vec R-\vec R_{0}))$  the average inter-particle distance 
$\langle R_{i}\rangle =\int d^{3}R_{i}W$ after release at $t=0$ obeys the relation
\begin{equation}
\label{e35}
\frac{\partial}{\partial t}\langle R_{i}\rangle =\zeta^{-1}_{ij}F^{ef}_{j}
\end{equation}                 
where the effective force is
\begin{equation}
\label{e36}
F^{ef}_{i}=-\frac{\partial U_{tot}}{\partial R_{i}}+T\zeta_{ip}\hspace{0.1cm}\frac{\partial}{\partial R_{q}}\hspace{0.1cm}\zeta^{-1}_{pq}
\end{equation}
This force consists of the potential part and the noise-induced drift, it is measured in the optical tweezers instant 
experiments in contrast to the long-time-statistics experiments, which give only the first potential term. Under 
experimental conditions \cite{GRIER2} the noise-induced drift is small. But if two particles are electrostatically fixed
in the middle plane between two glass plates, separated by the length $h$ ($R\ll h$), the effect of the noise-induced 
drift on the initial motion is not small
\begin{equation}
\label{e37}
F^{ef}_{i}=-\frac{\partial}{\partial R_{i}}\left[U_{tot}(R)+T\hspace{0.1cm}\frac{3a}{4R}\right]
\end{equation}
This equation is valid in the limit of small particle radius comparing to $R$. As one can see from Eq.~(\ref{e37}), optical
tweezers measurements can produce a deviation from a real interaction potential.     
\section{DISCUSSION} 
As shown in this paper, for hard spheres (no mean-field interaction excepting an infinite repulsion on contact) in a fluid
there are two types of hydrodynamic fluctuation forces: (i) the van der Waals forces, mediated by fluctuations of sound
waves ($U_{vdW}$), which is similar to the conventional (electromagnetic) van der Waals interaction, and (ii) forces due 
to a thermal drift of particles to the region with a bigger effective mass ($I$). Despite the van der Waals interaction 
(\ref{e8}) formally coincides with the electromagnetic van der Waals formula at $\varepsilon\rightarrow\infty$ \cite{DZYAL}, 
existence of such effect in hydrodynamics is not trivial, since a calculation on base of the hydrodynamic stress tensor 
leads to an incorrect conclusion of absence of the effect (Section III). The interaction $I$ is also of a fluctuation 
origin. It has a simple mechanical explanation based on that a particle spends more time at a region with a bigger mass, 
since there its velocity is smaller. The proposed hydrodynamic fluctuation interactions provide a long range (micron scale)
attraction of the order of $kT$ in contrast to the conventional van der Waals attraction, which is negligible at the micron
scale.   

Charged colloidal particles in an electrolyte, generally, cannot be considered as hard spheres, nevertheless, the Coulomb 
effects modify $U_{vdW}$ weakly, as shown in Section II. The interaction $I$ is not modified by the Coulomb effects in a 
bulk fluid, but can be influenced by an electrostatic restriction of fluctuations in the direction perpendicular to 
charged confining plates.

\begin{figure}[p]
\begin{center}
\vspace{1.5cm}
\epsfxsize=\hsize
\epsfxsize=13cm
\leavevmode
\epsfbox{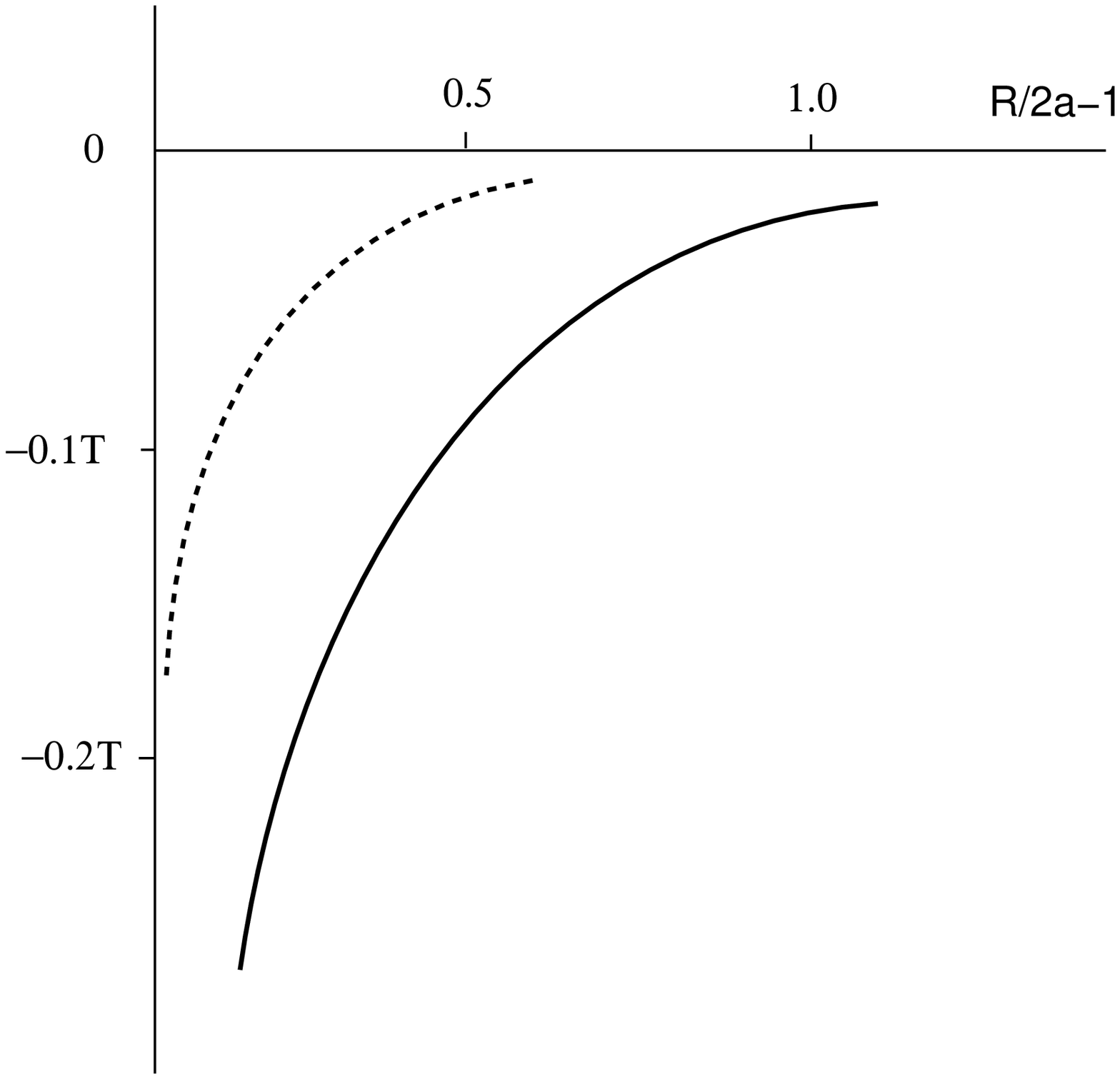}
\vspace{2cm}
\caption{The fluctuation interactions of two cylindrical particles with the axis of the diameter length directed 
perpendicular to two parallel plates, between which they are confined. The inter-plate distance is of the particle 
diameter. The dashed line is $I(R)$ and the solid line is the total fluctuation hydrodynamic interaction 
$U_{vdW}(R)+I(R)$.}
\label{fig2}
\end{center}
\end{figure}

The total interaction energy of two particles in a fluid is a sum of the Coulomb repulsive part (DLVO) \cite{DER,ISR} and 
the attractive potentials $U_{vdW}$ and $I$, plotted in Fig.~\ref{fig2}. A character of the resulting interaction depends on
particle charges and the Debye length in a very delicate way. Formally, the power law fluctuation attractions always win
the exponential repulsion at some distance $R_{0}$. Nevertheless, when $\lambda_{D}$ is not sufficiently small, the resulting 
potential minimum is far and can be indistinguishable in experiments. Refs.~\cite{GRIER5,GRIER6} reported an absence of 
attraction. A lucky choice of electrostatic parameters to observe fluctuation hydrodynamic forces corresponds to a close 
position of the minimum $(R_{0}-2a)\lesssim 2a$.

The hydrodynamic fluctuation interactions is unavoidable and therefore non-interacting spheres, strongly speaking, cannot 
exist in a fluid. The term ``hard spheres'' means an absence of a mean-field (non-fluctuation) interaction excepting an
infinite repulsion on contact. An interaction between two hard spheres in a dense system of them 
\cite{PUS,GAS,MEG,PHA,ZHU,CHE}, remaining of the order of $kT$, is expected to differ from interaction between two isolated
ones, considered here, since the hydrodynamic fluctuation forces, due to their non-perturbative nature, cannot be reduced 
to a pairwise interaction. 

The experimentally observed attraction satisfies the $k_{B}T$-universality condition, i.e., it is of the order of $k_{B}T$ in
various experiments. This universality is not a trivial property, since electrostatic and hydrodynamic mean-field energies
are of a few orders of magnitude bigger comparing to the room temperature. The $kT$-universality enables a strong 
selection of possible attraction mechanisms. Perhaps, a unique candidacy, satisfying this universality, is an attraction, 
mediated by thermal fluctuations of some physical quantities. Since the both mechanism proposed above lead to attraction 
of the order of $k_{B}T$, they are likely relevant in interpretation of experimental data on attraction. The next step is 
to study $U_{vdW}$ and $I$ at all $R$ for spherical particles in a bulk fluid and for a confined geometry. A contribution of
fluctuating surface waves to formation of attraction of particles on liquid-air \cite{JAIME2} and liquid-liquid \cite{NIKOL}
interfaces should increase the attraction effect.
\begin{center}
{\bf ACKNOWLEDGMENTS}
\end{center}
I would like to thank C.-K. Au, P. Chaikin, A. Ladd, M. Medina-Noyola, P. Pincus, R. Rajagopalan, J. Ruiz-Garcia, and 
N. G. van Kampen for valuable discussions.

\newpage

\newpage


\begin{references}

\bibitem{DER}

B .V .Derjagin, {\it Theory of Stability of Colloids and Thin Films} (Consultants Bureau, New York, 1989)

\bibitem{ISR}

J. Israelachvily, {\it Intermolecular and Surface Forces} (Academic Press, San Diego, 1991)

\bibitem{RAJ}

P. C. Heimenz and R. Rajagopalan, {\it Principles of Colloid and Chemistry} (Marcel Dekker, New York, 1997)

\bibitem{PIER}

P. Pieranski, Phys.Rev.Lett. {\bf 45}, 569 (1980)

\bibitem{ONODA}

G. Y. Onoda, Phys.Rev.Lett. {\bf 55}, 226 (1985)

\bibitem{ISE}

K. Ito, H. Yoshida, and N. Ise, Science {\bf 263}, 66 (1994)

\bibitem{JAIME1} 

J. Ruiz-Garcia, R. Gamez-Gomez, and B. I. Ivlev, Physica A {\bf 236}, 97 (1997)

\bibitem{JAIME2}

J. Ruiz-Garcia, R. Gamez-Corrales, and B. I. Ivlev, Phys.Rev. E {\bf 58}, 660 (1998)

\bibitem{JAIME3}

J. Ruiz-Garcia and B. I. Ivlev, Molec.Phys. {\bf 95}, 371 (1998)

\bibitem{JAIME4}

S. J. Mejia-Rosales, R. Gamez-Gomez, B. I. Ivlev, and J. Ruiz-Garcia, Physica A {\bf 276}, 30 (2000)

\bibitem{PINCUS}

P. M. Chaikin, P. Pincus, S. Alexander, and D. Hone, J.Colloid and Interface Sci. {\bf 89}, 555 (1982)

\bibitem{SINHA}

E. B. Sirota, H. D. Ou-Yang, S. K. Sinha, P. M. Chaikin, J. D. Axe, and Y. Fujii, Phys.Rev.Lett. {\bf 62}, 1524 (1989)

\bibitem {GAST}

Y. Monovoukas and A. P. Gast, J.Colloid and Interface Sci. {\bf 128}, 533 (1989)

\bibitem{LARSEN}

A. E. Larsen and D. G. Grier, Phys.Rev.Lett. {\bf 76}, 3862 (1996)

\bibitem{RADZ}

L. Radzihovsky, E. Frey, and D. Nelson, Bull.Am.Phys.Soc. {\bf 45}, 698 (2000)

\bibitem{LIN}

K.-H. Lin, J. C. Crocker, and A. G. Yodh, Bull.Am.Phys.Soc. {\bf 45}, 698 (2000)

\bibitem{SOW}

C.-H. Sow, C. A. Murray, R. W. Zehner, and T. S. Sullivan, Bull.Am.Phys.Soc. {\bf45}, 699 (2000)

\bibitem{MURRAY}

C. A. Murray and D. H. Van Winkle, Phys.Rev.Lett {\bf 58}, 1200 (1987)

\bibitem{ZAHN}

K. Zahn, R. Lenke, and G. Marett, Phys.Rev.Lett, {\bf 82}, 2721 (1999)

\bibitem{NEL}

T. Chou and D. R. Nelson, Phys.Rev.E {\bf 48}, 4611 (1993)

\bibitem{RUBI1}

J. Rubi and J. M. G. Vilar, J.Physics: Cond.Matt. {\bf 12}, A75 (2000)

\bibitem{MADDEN}

H. Lowen, P. A. Madden, and J.-P. Hansen, Phys.Rev.Lett. {\bf 68}, 1081 (1992)

\bibitem{OSPECK}

M. Ospeck and S. Fraden, J.Chem.Phys. {\bf 109}, 9166 (1998)

\bibitem{TATA}

B. V. R. Tata, M. Rajalakshmi, and A. K. Arora, Phys.Rev.Lett. {\bf 26}, 3778 (1992)

\bibitem{PALBERG}

T. Palberg and M. Wurth, Phys.Rev.Lett. {\bf 72}, 786 (1994)

\bibitem{FRAD}

G. M. Kepler and S. Fraden, Phys.Rev.Lett. {\bf 73}, 356 (1994)

\bibitem{ARAUZ}

M. D. Carbajal-Tinoco, F. Casro-Roman, and J. L. Arauz-Lara, Phys.Rev. E {\bf 53}, 3745 (1996) 

\bibitem{GRIER1}

J. C. Crocker and D. G. Grier, Phys.Rev.Lett. {\bf 77}, 1897 (1996)

\bibitem{GRIER2}

D. G. Grier, Nature {\bf 393}, 621 (1998)

\bibitem{GRIER3}

A. E. Larsen and D. G. Grier, Nature {\bf 385}, 230 (1997)

\bibitem{GRIER4}

D. G. Grier, Current Opinion in Colloid and Interface Science, {\bf 2}, 264 (1997)

\bibitem{JAIME5}

J. Ruiz-Garcia, private communication

\bibitem{NIKOL}

M. Nicolaides, PhD thisis, Technische Universit\"{a}t, M\"{u}nchen, 2001; M. Nicolaides, private communication.

\bibitem{SOGAMI}

I. Sogami and N. Ise, J.Chem.Phys. {\bf 81}, 6320 (1984)

\bibitem{BOWEN}

W. R. Bowen and A. O. Sharif, Nature {\bf 393}, 663 (1998)

\bibitem{NEU}

J. Neu, Phys.Rev.Lett. {\bf 82}, 1072 (1999)

\bibitem{SADER}

J. E. Sader and D. Y. C. Chan, J.Colloid and Interface Sci. {\bf 213}, 268 (1999)

\bibitem{STAT}

L. D. Landau and E. M. Lifshitz, {\it Statistical Physics}, Part 2, Butterworth-Heinemann, 1996

\bibitem{NIN1}

J. Mahanty and B. W. Ninham, {\it Dispersion Forces} (Academic Press, 1976)

\bibitem{NIN2}

B. W. Ninham and J. Daicic, Phys.Rev.A {\bf 57}, 1870 {1998}

\bibitem{NIN3}

H. Wennerstr\"{o}m, J. Daicic, and B. W. Ninham, Phys.Rev.A {\bf 60}, 2581 (1999)

\bibitem{SURESH}

L. Suresh and J. Y. Waltz, J.Coll. Interface Sci. {\bf 196}, 177 (1999)

\bibitem{BEVAN}

M. A. Bevan and D. C. Prieve, Langmuir {\bf 15}, 7925 (1999)

\bibitem{KIEFER}

J. E. Kiefer, V. A. Pasegian, and G. H. Weiss, J.Coll.Interface Sci. {\bf 51}, 543 (1975)

\bibitem{CZARNECKI1}

J. Czarnecki and T. Dabros, J.Coll.Interface.Sci. {\bf 78}, 25 (1980)

\bibitem{CZARNECKI2}

J. Czarnecki and V. Itschenskij, J.Coll.Interface Sci. {\bf 98}, 590 (1984)

\bibitem{PRIEVE}

D. C. Prieve and W. B. Russel, J.Coll.Interface Sci. {\bf 125}, 1 (1988)

\bibitem{AU}

C.-K. Au, J.Phys. B: At.Mol.Opt.Phys. {\bf 31}, L699 (1998)

\bibitem{TOKU}

M. Tokuyama, Phys.Rev.E {\bf 59}, R2550 (1999)

\bibitem{MANN}

J. Ray ans G. Manning, Langmuir {\bf 10}, 2450 (1994)

\bibitem{BRUIN}

N. Gronbech-Jensen, R. J. Mashl, R. F. Bruinsma, and W. M.Gelbart, Phys.Rev.Lett. {\bf 78}, 2477 (1997)

\bibitem{LOWEN}

E. Allahyarov, I. D'Amico, and H. Lowen, Phys.Rev.Lett. {\bf 81}, 1334 (1998)

\bibitem{SHKL1}

B. I. Shklovskii, Phys.Rev.E {\bf 60}, 5802 (1999)

\bibitem{SHKL2}

T. T. Nguyen, A. Yu. Grosberg, and B. I. Snklovskii, J.Chem.Phys. {\bf 113}, 1110 (2000)

\bibitem{LAU}

A. W. Lau, D. Levine, and P. Pincus, preprint cond-mat/0006266 (2000)

\bibitem{MESSINA}

R. Messina, C. Holm, and K. Kremer, Phys.Rev.Lett. {\bf 85}, 872 (2000)

\bibitem{TRIZAC}

E. Trizac, preprint cond-matt/0006069 (2000)

\bibitem{PARKER}

J. L. Parker, P. Richetti, and P. Kelichev, Phys.Rev.Lett. {\bf 68}, 1955 (1992)

\bibitem{KAPLAN}

P. D. Kaplan, L. P. Faucheux, and A. J. Libchaber, Phys.Rev.lett. {\bf 73}, 2793 {1994}

\bibitem{GOET}

B. G\"{o}tzelman, R. Evans, and S. Dietrich, Phys.Rev.E {\bf 57}, 6785 (1998)

\bibitem{VERMA}

R. Verma, J. C. Crocker, A. D. Dinsmore, and A. G. Yodh, Phys.Rev.lett. {\bf 81}, 4004 (1998)

\bibitem{METTEO}

J. C. Crocker, J. C. Metteo, A. D. Dinsmore, and A. G. Yodh, Phys.Rev.Lett. {\bf 82}, 4352 (1999)

\bibitem{MUD1}

T. M. Squires and M. P. Brenner, Phys.Rev.Lett. {\bf 85}, 4976 (2000)

\bibitem{COMM}

B. I. Ivlev, preprint cond-mat/0012212 (2000)

\bibitem{DZYAL}

I. E. Dzyaloshinskii, E. M. Lifshitz, and L. P. Pitaevskii, Adv.Phys. {\bf 10}, 165 (1961)

\bibitem{CAS}

H. B. G. Casimir and D. Polder, Phys.Rev. {\bf 73}, 360 (1948)

\bibitem{HYDR}

L. D. Landau and E. M. Lifshitz, {\it Fluid Mechanics}, Butterworth-Heinemann, 1997

\bibitem{LUB1}

W. Losert, L. Bocquet, T. C. Lubensky, and J. P. Gollub, Phys.Rev.Lett. {\bf 85}, 1428 (2000)

\bibitem{LUB2}

A. J. Levine and T. C. Lubensky, Phys.Rev.Lett. {\bf 85}, 1774 (2000)

\bibitem{MUD2}

R. B. Jones, Physica {\bf 105A}, 395 (1981)

\bibitem{MUD3}

D. Chan and L. White, Physica {\bf 122A}, 505 (1983)

\bibitem{CALDLEG}

A. O. Caldeira and A. J. Leggett, Ann.Phys. {\bf 149}, 374 (1983)

\bibitem{ITOSTRAT}

N. G. van Kampen, {\it Stochastic Processes in Physics and Chemistry}, North-Holland, Amsterdam, 1992 

\bibitem{STATPHYS1}

L. D. Landau and E. M. Lifshitz, {\it Statistical Physics}, Part I, Butterworth-Heinemann, Oxford, 1980

\bibitem{BLATTER}

G. Blatter and B. I. Ivlev, Phys.Rev. B {\bf50}, 10272 (1994)

\bibitem{MAZUR}

P. Mazur, Physica {\bf 110A}, 128 (1988)

\bibitem{LADD1}

A. J. C. Ladd, J.Chem.Phys. {\bf 88}, 5051 (1988)

\bibitem{LADD2}

A. J. C. Ladd, J.Chem.Phys. {\bf 90}, 1149 (1984)

\bibitem{LADD3}

A. J. C. Ladd, J.Chem.Phys. {\bf 93}, 3484 (1990)

\bibitem{ARAUZ1}

J. L. Arauz-Lara, private communication.

\bibitem{GRIER5}

J. C. Crocker and D. G. Grier, Phys.Rev.Lett. {\bf 73}, 352 (1994)

\bibitem{GRIER6}

S. H. Behrens and D. G. Grier, Bull.Am.Phys.Soc., {\bf46}, 179 (2001)

\bibitem{PUS}

P. N. Pusey, W. van Megen, P. Bartlet, B. J. Ackerson, J. C. Rarity,and S. M. Underwood, Phys.Rev.Lett. {\bf 63}, 2753 
(1989)

\bibitem{GAS}

A. Gast and Y. Monovouskas, Nature, {\bf 351}, 553 (1991)

\bibitem{MEG}

W. van Megen and S. M. Underwood, Phys.Rev.Lett. {\bf 70}, 2766 (1993)

\bibitem{PHA}

S.-E. Phan, W. B. Russel, Z. Cheng, J. Zhu, P. M. Chaikin, J. H. Dunsmuir, and R. H. Ottewill, Phys.Rev.E {\bf 54}, 6633 
(1996)

\bibitem{ZHU}

J. Zhu, M. Li, R. Rigers, W. Meyer, R. H. Ottewill, STS-73 Space Shuttle Crew, W. B. Russel, and P. M. Chaikin, Nature,
{\bf 387}, (1997)

\bibitem{CHE}

Z. Cheng, W. B. Russel, and P. M. Chaikin, Nature, {\bf 401}, 893 (1999)


\end{references}
\end{document}